# Models and Techniques for Ensuring Reliability, Safety, Availability and Security of Large Scale Distributed Systems

Valentin Cristea*, Ciprian Dobre*, Florin Pop*, Corina Stratan*, Alexandru Costan*, Catalin Leordeanu*, Eliana Tirsa*

*Faculty of Automatic Control and Computers, University Politehnica of Bucharest
( e-mail: valentin.cristea, ciprian.dobre, florin.pop, corina.stratan, alexandru.costan, catalin.leordeanu, eliana.tirsa @ cs.pub.ro).

**Abstract:** This paper presents an original approach to the development of models, methods and techniques for increasing reliability, availability, safety and security of large scale distributed systems, particularly Grids and Web-based distributed systems. The characteristics of these systems pose problems to ensuring dependability, especially because of the geographical distribution of resources and users, volatility of resources that are available only for limited amounts of time, and constraints imposed by the applications and resource owners. The problems of ensuring dependability represents a hot research subject today and, despite the fact that many projects obtained valuable results in this domain, no acceptable solution was yet found that could integrate all the requirements for designing a dependable system and that could exploit all the capabilities of modern systems. The paper proposes the design of an architectural model that allows the unitary and aggregate approach to reliability, availability, safety and security.

## 1. INTRODUCTION

Both in the academic and industrial environments there is an increasing interest in large scale distributed systems, which currently represent the preferred instruments for developing a wide range of new applications. The Grid computing domain has especially progressed during the last years due to the technological opportunities that it offers. While until recently the research in the distributed systems domain has mainly targeted the development of functional infrastructures, today researchers understand that many applications, especially the commercial ones, have some complementary necessities that the „traditional" distributed systems do not satisfy.

Together with the extension of the application domains, new requirements have emerged for large scale distributed systems; among these requirements, reliability, safety, availability, security and maintainability (synthesized in the literature in the concept of „dependability" (Avizenis et al.,2001)) are needed by more and more modern distributed applications, not only by the critical ones.

Although the importance of dependable systems is widely recognized and many research projects have been initiated recently in this domain, there are no mature implementations of these concepts available yet; the existing systems offer only partial solutions, and often the approaches separate the issues of reliability, availability, security etc.

In this paper we present innovative solutions to researching and developing original models, methods and techniques that satisfy dependability requirements in large scale distributed systems, which are essential for the new types of applications. We present an architectural model which allows the unified approach to reliability, availability, safety and security. The model is based on the use of tools for communication, monitoring, scheduling, management and accessibility (retrieval, efficient transfer) of data, and security, in order to provide increased levels of dependability. We also present a minimal set of functionalities, absolutely necessary to ensure the reliability, availability, safety and security. The presented architecture is adaptable, reconfigurable and regenerative and can help distributed system designers to increase dependability while preserving scalability.

The rest of this paper is organized as follows. Section 2 describes the requirements of implementing a complete dependable solution. In Section 3 we present related work. Section 4 presents the architectural design being proposed in the current paper. In Section 5 we conclude and present future work.

## 2. REQUIREMENTS

The characteristics of large scale distributed systems make dependability a difficult problem from several points of view. A first aspect is the geographical distribution of resources and users that implies frequent remote operations and data transfers; these lead to a decrease in the system's safety and reliability and make it more vulnerable from the security point of view. Another problem relates to the volatility of the resources, which are usually available only for limited periods of time; the system must ensure the correct and complete execution of the applications even when resources are introduced and removed dynamically, or when they are damaged. The management of distributed systems is also complicated by the constraints that the applications and the owners of the resources impose; in many cases there are conflicts between these constraints – for example, an application needs a long execution time and performs database operations, while the owner of the machine on which the application could be run only makes it available in a restricted time interval and does not allow database operations.

Solving these issues still represents a research domain and, although numerous projects have obtained significant results, no complete solution has been fond yet to integrate all requirements involved in obtaining a dependable system.

In order to facilitate implementation, as well as evaluation, the architectural model being proposed provides trust mechanisms for service and resource provisioning, and guarantees for the delegation and context services, together with various mechanisms to orchestrate services such that to achieve high levels of reliability and fault tolerance.

Unlike similar research solutions previously presented, we propose a unified approach to all the aspects concerning dependable systems, by analyzing and designing methods and techniques for improving the reliability, availability, safety and security of large scale distributed systems. We also explore issues that appear in reference publications, but today are insufficiently addressed. For example, the fault detection and correction mechanisms are extended with fault prediction and avoidance mechanisms. The architectural model proposed in this paper also includes a prediction component, and its design is based on the use of monitoring tools.

We are focusing on Grid systems, but the proposed solutions can be easily applied to other types of distributed systems, for example Web based ones.

Contemporary distributed environments are composed of multiple autonomous services, possible running on different machines, under the control of different organizations. Since our focus is on Grid systems we also need to take into account the concept of Virtual Organizations. From a security point of view this poses a number of interesting challenges. If we view the Virtual Organizations as groups of users who share a certain level of trust we need to create security policies which are active within a VO and others between members of a VO and non-members. For example, a member of a VO might allow other members to read data he has submitted but there must be mechanisms in place to ensure that this is not possible for a user which is not a part of the VO. At the same time the level of trust between members of a VO is not absolute so in this case we must also make sure that users cannot modify each other's data.

The scalability and availability requirements of such services have led to system architectures that diverge significantly from those of traditional distributed systems. Given the need for thousands of nodes, cost often necessitates the use of inexpensive personal computers wherever possible, and efficiency often requires customized service software. Likewise, addressing the goal of zero downtime requires human operator involvement and pervasive redundancy within clusters and between globally distributed computing centers. The new requirements for unrestricted availability demands coming from next-generation critical applications pose interesting challenges because the large scale distributed systems:

- typically comprise many inexpensive PCs that lack expensive reliability features;
- undergo frequent scaling, reconfiguration, and functionality changes;
- often run custom software that has undergone limited testing;
- rely on networks within service clusters, between geographically distributed collocation facilities, and between collocation facilities and end-user clients;
- aim to be available 24/7 for users worldwide, making planned downtime undesirable or impossible.

The proposed architectural model must, therefore, be instrumented with the support for fault-tolerance and security techniques to assure the resiliency of applications and the high-availability of crucial Grid services. Among the requirements that must be satisfied by the architecture we recall tamper resistance in the face of intruders, the system must always be invoked and must be designed small enough such that to be subject to analysis and tests to ensure its correctness using formal methods.

## 3. RELATED WORK

Most of the approaches to the dependability problem are mainly focused on fault detection and fault tolerance. The authors of (Jin et al.,2006) introduce an adaptive system for fault detection in Grids and a policy-based recovery mechanism. Fault detection is achieved by monitoring the system, and for recovery several mechanisms are available, among which task replication and checkpointing. In (Sommerville et al.,2003) the authors present a solution for web services, based on the implementation of a fault-tolerant container; the fault-tolerant containers manage a set of replicated services, which can be offered by external providers as well. The containers can be configured for various fault tolerance strategies (either an equivalent service is invoked when a service fails, or multiple equivalent services are invoked from the beginning and a voting mechanism is applied etc.).

Another approach less related to large scale distributed systems is presented in (Cox et al., 2005), which addresses the tolerance to hardware faults through virtualization; the proposed solution is named Loosely Synchronized Redundant Virtual Machines (LSRVM).

In (Grimshaw et al., 2005) the authors emphasize on the idea of survivability, which consists in providing, in case of a error, an alternative service, even if the latter doesn't fully satisfy the initial user requirements. For requirements specification, the authors propose DESL (Dependability Exchange and Specification Language), an XML based language; DESL isn't fully specified yet.

Other two papers in this field are (Sonnek et al, 2005), which proposes a reputation based classification of Grid services, similar to the one in peer-to-peer systems, and (Rilling et al, 2006), which presents a Grid operating system architecture, with self-healing properties. In (Neocleous et al., 2006) the authors carry out a detailed study on errors in Grid systems, providing case-studies from the EGEE (Enabling Grids for E-Science) European project.

Another aspect of dependability is security. Currently there are several solutions for providing security for Grid environments and for distributed systems in general. In case of Grid systems, in (Foster et al, 1998) the authors identified a set of base requirements (single authentication, credentials protection, interoperability with local security solutions, etc) and proposed both an architectural model for security and a public key cryptography infrastructure based on a reference implementation (Grid Security Infrastructure – GSI).

The existing solutions for security in large scale distributed systems have various unresolved issues (Arenas et al, 2006). The majority of Grid systems were initially designed for scientific collaboration between participants that knew each others. This implies an implicit trusting relationship, all partners sharing a common goal – for example to carry out a scientific experiment – and it is implicitly assumed that the resources are provided and shared according to some well defined and respected rules. But when used in industrial environments such systems must satisfy the need to share resources with unknown groups, which could generate certain risks. A possible solution is to impose certain mechanisms to preserve the identity even in this situation.

## 4. ARCHITECTURAL MODEL

In our proposal we aim to exploit some features of the services provided by the large-scale distributed systems to enhance availability:

- Plentiful hardware allows for redundancy.

- Geographic distribution of collocation facilities allows control of environmental conditions and resilience to large-scale disasters.

The model considers that the problem of developing highly dependable large scale distributed systems is technological prohibited by the various technical problems arising in distributed systems, such as the weak resilience of the operating systems, known for offering various security vulnerabilities. The solution to this problem consists in the inclusion of a complete set of mechanisms necessary to guarantee the various security needs: the trust in service and resource provisioning, guarantees to the delegation and context mechanisms. The architectural model includes for this purpose two specialized components.

A first set of components act inside the middleware layer of the distributed system, intermediating dependability between different hosts composing the distributed systems. At this level the components of the proposed architectural model are presented in Figure 1.

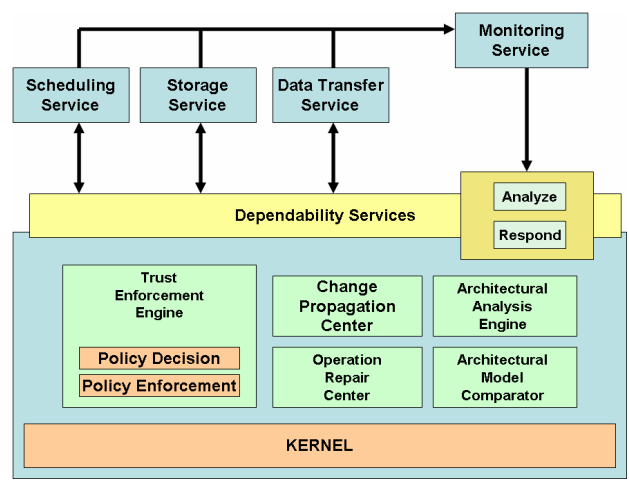

fig. 1. The architectural model of the service layer

At the bottom of the architecture is the core of the system designed to orchestrate the functionalities provided by the other components. Its role is to integrate and provide a dependable execution environment for several other components. It also orchestrates the survivable and redundancy mechanisms described below.

The model is based on a minimal set of functionalities, absolutely necessary to ensure the reliability, availability, safety and security. We argue that such functionalities are a reference for implementing a functional core for a dependable distributed system and that the set of functionalities are sufficient to cover the basic needs of the majority of distributed systems, without operating major changes.

The architecture includes the mechanisms necessary to ensure a strong resilience in the presence of faults and various security threats coming from the incoming connection. At this level we included a first component based with detecting when various processes of the distributed system fail. We propose an approach based on adaptive, decentralized failure detectors, capable of working asynchronous and independent on the application flow. The proposed solution considers architecture for the failure detectors, based on clustering, the use of a gossip-based algorithm for detection at local level and the use of a hierarchical structure among clusters of detectors along which traffic is channelled. The solution can scale to a large number of nodes, considers the QoS requirements of both applications and resources, and includes fault tolerance and system orchestration mechanisms, added in order to assess the reliability and availability of distributed systems.

At this level we extend the solution of the inclusion of a complete set of mechanisms necessary to guarantee the various security needs previously described with the mechanisms of taking security related decisions, possible based on monitored data. A component also integrated into the architecture is responsible with ensuring that these decisions are enforced by its integration with the various existing services and security infrastructures (such as GSI or Kerberos).

Beside the traditional alternatives to constructing dependable services as were integrated here, the fault avoidance, fault elimination and fault tolerance (Nguyen-Tuong et al, 2004), we also considered mechanisms for the dynamic adaptation to environment conditions. For this reason the model includes a minimal set of components: a monitoring subsystem, a component used to compare various architectural patterns, an architectural core for analyzing data, a component responsible with starting adequate repairs and a module responsible with the broadcast of information regarding the changes occurring in various part of the distributed system. The architectural model thus offers various intelligent mechanisms designed to dynamically solve the various occurring problems and that also represents a link between the distributed system and the various running functionalities of the applications, using automatic reconfiguration of both services and resources.

The proposed architectural model integrates several technologies that are necessary to ensure a high level of reliability, providing fault prevention, fault removal, fault tolerance, and fault forecasting, as well as security mechanisms.

The core allows modularity, an aspect which influences the adoption of the architecture in various existing Grid infrastructures. The modularity allows dealing with the service-oriented architecture on which existing grid systems are built and facilitates the implementation of the survivability concept presented below. The model also allows the generic integration with various technologies specific to large scale distributed systems, and the development and evaluation of various new models and technologies. For this, the proposed components ensure the interoperability with services specific to such systems and each one ensures the reliability specifications coming from higher-level applications.

Another problem with existing distributed system relates to the prohibitive costs imposed by the replication of all basic components of the system, as well as those responsible for the communication, storage and computation. Replication of these services is currently seen as a necessity because it offers the highest availability to the system. In fact, several of the previously analyzed solutions are based on the idea of replicating as many services and resources as possible such that, when one fails, the other immediately takes its place and functionality. The use of replication mechanisms is appealing because it increases the reliability of the system, but our model considers a combination of both replication mechanisms together with mechanisms to ensure survivability of the system in the presence of major faults.

The adopted solution to developing an architectural model in which the system survive by adapting in the presence of fault arise naturally by explicitly acknowledge the impossibility to include a complete solution to ensure reliability considering the resulting resources and technologies. The impossibility is mainly derived from the human nature of the system developers and operators, as well as prohibitive costs necessary to implement a completely replicated Grid system. For such reasons we adopt the strategy of using replication only for the most basic core functionality of the system. We use replication in the form of fault-tolerant containers; the fault-tolerant containers could easily manage a set of replicated services. We also adopt a survivability approach in which we provide several services such that when one can not tolerate faults or security attacks another one takes its place. The system survives by learning the conditions which eventually led to the fault of a service or resource, using patterns.

The architectural model also includes mechanisms necessary for the system to adapt and survive in the context of dynamic nature of existing resources, faults occurring in the system and various changes in the exploitation requirements of the system. In order to offer technologies capable to automatically correct various occurring problems and that are capable of taking intelligent decisions we propose the adoption of a solution based on the use of architectural patterns for correcting faults. Such patterns have a special role in the context of monitoring, being used in a process of automatically detect faulty behavior of resources or various other problems and take higher-level decisions to start adequate repairing steps (Garlan et al, 2003).

For that, the Analyze and Respond services (Figure 2) would continuously analyze real-time data collected from an external monitoring system. The data are processed and compared against a set of patterns (existing and learned as the system adapts to changing environment). One of the biggest challenges at this level is represented by the capability to not only recognize faults, but also predict faulty behavior in various parts of the distributed system. This service is in fact capable to do that by using various mathematical distributions and AI constructs (neural networks, moving average distributions, etc.).

The component responsible with the real-time analysis of monitored data is the Architectural Analysis Engine (see Figure 1). The analyzer uses the services provided by the Architectural Model Comparator. When faults or security breaches are detected the system would further initiate (possible automatically, if not possible would alert a human operator of the problem) an operation to correct the situation. Such a correction is initiated by the Operation Repair Center component (Figure 1). The correction could be, for instance, the raise of an alternative service that could supplement the activity offered by the faulty service or could represent an instruction sent to a Scheduling Service to reschedule faulty jobs. Also, by integrating various existing services, if the underlying distributed environment is capable of executing checkpointing operations (such as Condor) then the Operation Repair Center could also initiate an action to recover the last state using the checkpointing data.

Of course, once the repair action was determined it must be correctly propagated inside the entire distributed system. The role of this action is assumed by the Change Propagation Center (Figure 1). In this way, the architecture involved communication with services providing functionalities in various parts of a Grid system. The data is collected from a monitoring service and we intend to integrate the architecture closely with such a system. The checkpointing capability involves dealing with a storage service. The state of the system must be received from a central indexing service of the underlying middleware. The fault of a job would initiate a dialog between the services provided by the presented architecture and a scheduling service. The detection of faults, as well as propagation of recovery actions means using a data transfer service.

An approach based on offering survivable architectural characteristics naturally maps on a developing environment in which applications are built by incorporating existing elements in various administrative domains, an inherent characteristic of Grid systems. The specification of alternative services provides an environment in which existing elements, having different characteristics such as private politics to ensure resource availability, special requirements to system loading, private security politics, can be analyzed and easily incorporated without deforming the survivable character of the systems.

In such collaborative environments the reliability requirements of any application are handled by a process specialized in discovering the resources and that uses a composite set of services, having the role to provide the necessary levels of reliability.

A major challenge in the design of the presented approach consists in the investigation of the techniques and transformations necessary to enhance the reliability of the composite set of services, mostly by offering an architectural solution based on workflows.

The second architectural component set responsible with ensuring dependability at the operating system level is represented in Figure 2.

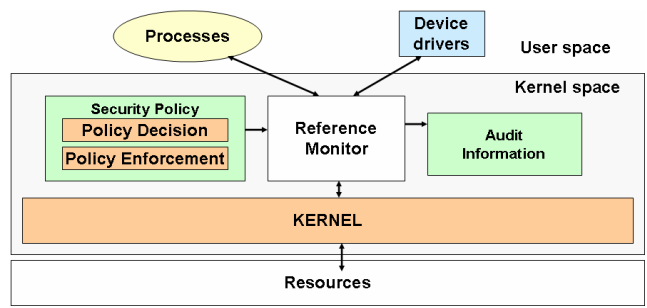

Fig. 2. The components acting at the kernel level

At the heart of the operating system lies the kernel, responsible with providing various services: signal and message passing services, memory and process management, etc. A first component is the Reference Monitor, which resides within the kernel space. This cannot be circumvented and/or modified and must be simple and compact enough to be readily understood. All requests coming from various processes or device drivers must go through this component, which basically acts as intermediary between the kernel and user-space applications. This is achieved by masking the fundamental system calls (for address-space management, inter-process communications and thread management) as being served by the kernel itself, but in reality all these calls are transferred to the Reference Monitor, are checked for eventual safety problems and, if everything is correct, further transferred to the kernel itself. In this way the Reference Monitor acts as a proxy for system calls being made between user-space applications and the kernel itself.

This component validates all attempts to access the system resources based on the input provided by the second component, the Security Policy. This component provides complete mediation of verification of the validity of requests, based on sets of permissions. All possible problems are reported and stored for further inspection and even for automatically pattern recognition of errors in future cases.

## 5. CONCLUSION

In this paper we presented an architectural approach to the development of models, methods and techniques for satisfying dependability requirements in large scale distributed systems. Dependability remains a key element in the context of application development and is by far one of the most important issues still not solved by recent research efforts.

Our work is concerned by increasing reliability, availability, safety and security, particularly in Grids and Web-based distributed systems. The characteristics of these systems pose problems to ensuring dependability, especially because of the heterogeneity and geographical distribution of resources and users, volatility of resources that are available only for limited amounts of time, and constraints imposed by the applications and resource owners. Therefore, we proposed the design of a hierarchical architectural model that allows a unitary and aggregate approach to dependability requirements while preserving scalability of large scale distributed systems.

Our original contribution addresses the architectural model, the core of the architecture, and the combination of existing monitoring, scheduling, data management, security and fault tolerance solutions to increase dependability of large scale distributed systems.